\documentclass[12pt]{article}
\usepackage{epsfig}
\def\remark#1{}

\newlength{\singlepicwidth} 
\newlength{\doublepicwidth} 
\newlength{\firstpicshift} 
\newlength{\secondpicshift} 
\def\remark#1{}

\newcommand{\pn}{\ensuremath{\mathrm{e}^+\mathrm{e}^-}}
\newcommand{\p}{\ensuremath{\mathrm{e}^+}}
\newcommand{\n}{\ensuremath{\mathrm{e^-}}}
\newcommand{\Ep}{\ensuremath{E_{e^+}}}
\newcommand{\En}{\ensuremath{E_{e^-}}}
\newcommand{\ES}{\ensuremath{E_{\Sigma}}}
\newcommand{\ED}{\ensuremath{E_{\Delta}}}
\newcommand{\Zr}{\ensuremath{^{90}\mathrm{Zr}}}
\newcommand{\Ox}{\ensuremath{^{16}\mathrm{O}}}
\newcommand{\Sr}{\ensuremath{^{90}\mathrm{Sr}}}
\newcommand{\Y}{\ensuremath{^{90}\mathrm{Y}}}

\newcommand{\Pbs}{\ensuremath{^{207}\mathrm{Pb}}}

\newcommand{\Bis}{\ensuremath{^{207}\mathrm{Bi}}}

\newcommand{\degree}{\ensuremath{\mathrm{^\circ}}}

\newcommand{\mucm}{\ensuremath{\mathrm{\,\mu g/cm^2}}}

\vspace*{2cm}
\begin{document}
\begin{center}
 
{\Large \bf \mbox{First Energy and Angle differential Measurements of}}
 
{\Large \bf \mbox{\pn -pairs emitted by Internal Pair Conversion of}} 

{\Large \bf \mbox{excited Heavy Nuclei}}

\end{center}

\vspace{3cm}
\noindent
{\large
 U.~Leinberger$^a$, E.~Berdermann$^a$, F.~Heine$^b$, S.~Heinz$^b$,
 O.~Joeres$^b$, P.~Kienle$^b$, I.~Koenig$^a$, W.~Koenig$^a$, 
 C.~Kozhuharov$^a$, A.~Schr\"oter$^a$, 
 H.~Tsertos$^{c,}$\footnote{Corresponding author, 
  e-mail "tsertos@alpha2.ns.ucy.ac.cy" \\
 Dept. Nat. Science, Univ. of Cyprus, PO 537, 1678 Nicosia, Cyprus}\\
 (ORANGE Collaboration at GSI)\\
 C.~Hofmann$^d$, G.~Soff$ ^d$
}
 \\[3mm]
 $^a$ Gesellschaft f\"ur Schwerionenforschung (GSI),
 D--64291 Darmstadt, Germany\newline
 $^b$ Technical University of Munich,  D--85748 Garching, Germany\newline
 $^c$ University of Cyprus, CY--1678 Nicosia, Cyprus\newline
 $^d$ Inst. f. Theoretische Physik, Technical University Dresden, D--01062 
      Dresden, Germany\newline 

\vspace{2.5cm}
\begin{center}
 (Revised version: December 1997)
\end{center}

\newpage
\begin{center}
 \section*{Abstract}
\end{center}       

We present the first energy and angle resolved measurements of \pn -pairs 
emitted from heavy nuclei ($Z\ge 40$) at rest by internal pair conversion 
(IPC) of 
transitions with energies of less than 2\,MeV as well as recent theoretical
results using the DWBA method, which takes full account of relativistic
effects, magnetic substates and finite size of the nucleus. The 1.76\,MeV E0
transition in  \Zr\ (\Sr\ source) and the 1.77\,MeV M1 transition in \Pbs\
(\Bis\ source) have been investigated experimentally using the essentially
improved setup at the double-ORANGE $\beta$-spectrometer of GSI. The
measurements prove the  capability of the setup to cleanly identify the IPC
pairs in the presence of five orders of magnitude higher $\beta^-$ and
$\gamma$ background from the same  source and to
 yield essentially
background-free sum spectra despite the large background. 
Using the ability 
of the ORANGE setup to directly determine the opening angle
of the \pn--pairs ($\Theta_{e^+e^-}$), the angular 
correlation of the emitted pairs was measured
within the range covered experimentally
($40^{\circ} \leq \Theta_{e^+e^-} \leq 180^{\circ}$).
In the \Zr\ case the 
correlation  could be deduced for a wide range of energy  differences \ED\ of
the pairs ($-530$ keV $\leq \ED \leq 530$ keV).
The \Zr\ results are in good agreement with  recent theory. The
angular correlation deduced for  the M1 transition in \Pbs\ is in strong
disagreement with theoretical  predictions derived within the Born
approximation and shows almost isotropic character. This is  again in
agreement with the new theoretical results.\\

{\bf PACS numbers:} 14.60.Cd, 23.20.En, 23.20.Ra, 29.25.Rm, 29.30.Aj

\newpage 
\section{Introduction}

Although known since the 1930's as one of the elementary decay channels of
exited nuclear states with transition energies of more than $2m_e c^2 =
1.022$\,MeV \cite{Opp33a,Jae35a}, only limited theoretical and very limited
experimental investigations of
 this process have been carried out in the
following years. 
Theoretical results concerning the angular correlation of the
\pn- pairs were mainly based on Born approximation
and thus valid only for point-like low-$Z$ nuclei and/or transition
energies much higher than the threshold of 1.022\,MeV
\cite{Dal51a,Hor54a,Ros49a,Hor43a,Opp41a,Ros35a}. Only in the early 1990's more
theoretical effort yielded results valid also for finite size nuclei and low
transition energies (below twice the  threshold) at least for electrical
multipole transitions \cite{Hof94a}.
Correct treatment of magnetic transitions was achieved only recently
\cite{Hof95a}, and the results for angular correlations are presented here.

Very few experimental results on IPC are published, and most of them
on total IPC coefficients only, derived from the integral \p\
yield \cite{Gre56a}.
However, one very early experiment \cite{Dev49a,Dev54a} already measured the
angular correlation  for the 6\,MeV E1 transition 
in \Ox,
but without any energy measurement. This and other transitions of even  higher
energies in light nuclei have been measured recently \cite{Fro95a} both energy 
and angle resolved. The results were in reasonable agreement with Born 
approximation, as expected, since this approximation should be valid for these
cases.
However, \pn -pairs emitted via IPC from heavy nuclei are of considerable 
interest, since IPC is the only known source of \pn-pairs with a sharp sum 
energy. 
Narrow lines with sum energies ($\ES = \Ep+\En$) in the range 
 $500\,keV \leq \ES \leq 800\,keV$
have been found in heavy-ion experiments around the Coulomb barrier at GSI
\cite{Cow86a,Ber88a,Kie88a,Kie89a,Koe89a,Bok90a,Sal90a}. 
The angular 
correlation of the emitted leptons is of particular interest, since the
response of the various setups depends strongly on it. Prior to this work, no
data have been available. Additionally, the most recent theoretical results
shown below should be tested experimentally.

\section{Theory of internal pair conversion}

The angular 
correlation of electron-positron pairs emitted by internal pair
conversion (multipolarity $L>0$) can be expressed in terms of Legendre
polynomials \cite{Hof96b}

\begin{equation}
\frac{{\rm d}^2\beta}{{\rm d}E \, {\rm d}\cos\vartheta} =
\frac{1}{2} \, \frac{{\rm d}\beta}{{\rm d}E} \left(
1 + \sum\limits_{i>0} a_i \, P_i( \cos\vartheta) \right).
\end{equation}

The differential pair conversion coefficient $\frac{{\rm d}\beta}{{\rm d}E}$
and the anisotropy coefficients $a_i$ are calculated numerically.
The calculation is performed using DWBA, i.e., the
electron and positron wave functions are taken to be the exact scattering
solutions of the Dirac equation for the Coulomb potential of an extended
nucleus \cite{Hof93a}. The differential pair conversion coefficient
and the anisotropy coefficients depend therefore on the nuclear charge 
number, as well as on the nuclear transition energy and the 
multipolarity and parity of the nuclear transition. Thus, the calculation
goes beyond the Born approximation, where electron and positron wave
functions are considered as plane waves \cite{Ros49a}.
$\frac{{\rm d}\beta}{{\rm d}E}$ is also tabulated in \cite{Hof95a}. 
Fig. 1 shows the angular correlation for M1 pair conversion of an
lead nucleus ($Z=82$) and a transition energy of $1.77$~MeV. The energy
difference $\ED = \Ep - \En$
amounts to $150$~keV, which corresponds to the experimental data
presented in this work (see Fig. 10 below).
The calculation was performed assuming a point-like
nucleus (dashed curve) as well as an extended nucleus (full curve). 
The point nucleus approximation is not very realistic for magnetic pair
conversion in highly-charged nuclei since it overestimates the pair emission
rate. Finally the dotted curve corresponds to the result obtained within
the plane wave Born approximation \cite{Ros49a}, which displays the typical
forward-peaked behaviour previously expected for the pair emission. 
For high nuclear charge numbers, the angular correlation deviates 
strongly from the Born approximation result. The pair emission 
occurs then nearly isotropic. This behaviour is elucidated
by Fig. 2, which depicts the angular correlation, assuming a $1.77$~MeV
M1 transition with symmetric splitting of the transition energy on the
electron and the positron and various nuclear charge numbers.

Also in the case of electric monopole pair conversion (denoted as E0 pair 
conversion) the angular correlation deviates from the Born approximation
result if one takes full account of the nuclear charge number
\cite{Hof96b}. The angular correlation is then written as
\begin{equation}
\frac{{\rm d}^2\eta}{{\rm d}E \, {\rm d}\cos\Theta} =
\frac{1}{2} \, \frac{{\rm d}\eta}{{\rm d}E} \left(
1 + \epsilon \, \cos\Theta \right).
\end{equation}
Again the differential pair conversion coefficient 
$\frac{{\rm d}\eta}{{\rm d}E}$ 
and the anisotropy coefficient $\epsilon$ are computed numerically by
employing the scattering solutions of the Dirac equation and assuming an
extended nucleus. Using these Coulomb-distorted plane waves  \cite{Hof93a}
internal pair conversion results in a strong dependence on the nuclear charge 
number. In Fig. 3 the anisotropy coefficient is plotted versus the kinetic
positron energy for the $1.76$~MeV E0 transition in Zr ($Z=40$) (full curve)
and U ($Z=92$) (dotted curve).  If the nuclear charge number is increased the
angular correlation starts to deviate from the Born approximation result
\cite{Opp33a} (dashed curve) and becomes more isotropic.

\section{Experimental setup}\label{setup}

As in the heavy-ion experiments \cite{Koe89a,Koe93a,Lei97a}, we use
two identical iron-free, orange-type $\beta$- spectrometers~\cite{Moll65},
facing each other with a common object point, at which a 
target wheel is placed (Fig. 4). 
Positrons (\p) emitted in the backward 
($\vartheta_{e^+} = 110^{\circ} - 145^{\circ}$) 
and electrons (\n) emitted in the forward
hemisphere ($\vartheta_{e^-} = 38^{\circ} - 70^{\circ}$)
are focussed onto the corresponding lepton detectors.
Each lepton  detector consists of an array of high-resolution 
Si PIN diodes, called PAGODA.
Intrinsic 
to this setup is the capability of focussing, at  
a given field setting, only a certain momentum interval of \p{} or \n{}
by rejecting the opposite charge.
This is of major advantage because the high $\beta^-$
background is suppressed completely on the \p--side, while
the selection of only a certain band of \n -momenta
on the \n--side enables operation with the strong sources required by the low 
branching ratios. 

The leptons are identified by matching their momentum, derived from the
hit-point on the PAGODA and the spectrometer field setting, with their energy,
measured by the PIN diodes. This efficiently suppresses backgrounds due to
$\gamma$ rays. An additional coincidence requirement with the 511~keV
annihilation radiation is thus not necessary for the \p{} identification. The
momentum-energy matching in addition rejects the $\approx 25\%$ leptons
backscattered from  the PIN diodes almost completely, a unique feature of this
setup.
As mentioned above, for the lepton identification only events for which 
the energy-momentum relation is fulfilled are accepted, with the result 
that the remaining lepton misidentification is
small and can be determined reliably. 
This is demonstrated in Fig. 5 for the case of the positron identification.
As can be seen a clear signature for the focussed positrons is obtained. 
In such a spectrum all other particles, like electrons and $\gamma$ rays
scattered from the vacuum vessel walls and the spectrometer coils as well
as positrons backscattered from the detectors, result in a broad continuous
background distribution which can be determined quantitatively.
The situation is very similar for the case of the electron identification.

Each of the two identical lepton detection systems consists of 72 segmented, 
high-resolution, 1 mm thick, Si- PIN 
diodes which are mounted on 12 roofs, 6 detectors on each roof (see Fig. 4).
Each detector chip has a trapezoidal shape
(24 mm base, 16 mm top, and 16 mm height) which is
tilted by $\approx 40^\circ$ relative to the spectrometer axis.
Each detector is also subdivided into three electrically separated 
sectors, each
covering an azimuthal angular range of $\Delta\phi_{lept}$ = $20^\circ$.
A matrix readout, each roof as well as each 
three neighboring sectors at the same azimuthal angle to one
preamplifier, records information on energy and arrival time of a lepton 
hitting a detector.                                                             
Cooling the detectors to $\approx -25^\circ$C, the e$^+$e$^-$-sum-energy 
and time resolution
achieved amounts to $\approx$15 keV
and $\approx$4 ns (FWHM), respectively.
       
The opening angle of the \pn--pair, $\Theta_{e^+e^-}$, is measured directly
within a range of
$40^{\circ}-180^{\circ}$ in the laboratory. Using the
$\phi$--separation of the PAGODA's, this range can be subdivided into 
ten angular bins (see table 1) with  
centroids in the range
$\Theta_{e^+e^-}$~=~70$^{\circ}$\dots 167$^{\circ}$. 
These angular bins cover the same solid angle of 0.62\,sr$^2$, except the first
and last, which cover only half of this value. This is due to simple 
combinatorics: for each of the angular bins all combinations between \p\ and 
\n\ PAGODA columns with the same $|\Delta\phi|$ are put together. For 
$|\Delta\phi|$=0\degree\ and 180\degree\ only one column on the other PAGODA 
contributes, but for e.g.\ $|\Delta\phi|$=20\degree{} both signs are 
possible and
two columns of the other PAGODA contribute to this $\Theta_{e^+e^-}$-bin.
The values given 
in table~1 have been calculated via a Monte Carlo simulation, which takes into
account  
a realistic source spot size and the shadowing of the coils 
via ray tracing. The same code, but extended to handle the angular 
correlation of the emitted pairs, small-angle scattering and energy loss 
in the source, was 
also used for results shown later in this paper.

Each pagoda array covers a maximum momentum acceptance of
$\Delta p/p$ =30\% which corresponds to an energy interval 
of $\Delta E \sim$150 keV  at a lepton energy of  300 keV.
Within this momentum interval, the full-energy peak efficiency 
is 10\% and 11\% of 4$\pi$, for electrons and positrons, respectively
(see also Ref. \cite{Lei97a}). 

To cover an area in the \ES-\ED\ plane larger than given by the momentum 
acceptance of the spectrometers, the spectrometer currents are stepped up and 
down in a correlated manner, where each step is done after the same live-time
interval. Thus an intrinsic correction for dead time effects due to different
count rates is achieved.

Under the assumption, that the $\theta$-dependant longitudinal dispersion 
of the spectrometers cancels, the
pair detection efficiency of the double-ORANGE setup can be written as
follows:

\begin{equation}
\varepsilon=C_{spec}(E_\Sigma,E_\Delta,I_{spec}) \times
C_{angle}(\Theta_{e^+e^-}),
\end{equation}
where $C_{spec}$ is the detection efficiency of the spectrometers  for 
isotropic emission, depending on energy distribution and spectrometer current
sweep, 
and $C_{angle}$ depends on the angular correlation of the leptons.
However, for any given angular correlation, the relative number of counts
in every angular bin of the setup does not depend on energy 
distribution or spectrometer fields. The angular correlation can thus be
deduced from the relative intensities alone. The above mentioned assumption 
is justified, if leptons entering the spectrometers at any accepted angle 
$\theta$ from the axis and passing the spectrometer will hit any of the 
pagoda's detectors.
For a given, fixed spectrometer current this is true 
for the flat top of the efficiency of $\frac{\Delta p}{p}\approx30\%$. At 
the edges of the flat top the efficiency drops fast to zero, and the 
efficiency in this case depends  on  $\theta$.
When the spectrometers are used in current sweep mode, the flat top region 
is extended over the whole acceptance range covered, and a $\theta$ 
dependance is present only at the steep edges.

\section{Experimental results}

With the setup described above we investigated the 1.76\,MeV E0 transition in 
\Zr\ and the 1.77\,MeV M1 transition in \Pbs, using radioactive sources.

\subsection{The 1.76\,MeV E0 transition in \Zr}

The first excited $0^+$ state at 1.76\,MeV of the $0^+$ nucleus \Zr\ is 
forbidden for normal $\gamma$ decay and it's main decay channels are internal 
conversion, emitting electrons with sharp energy given by the transition 
energy minus the \n\ binding energy ($\approx2/3$) and internal pair 
conversion ($\approx1/3$)\footnote{
A very small part actually 
decays via multiple $\gamma$ emission. The
strongest of these branches emits two $\gamma$'s back to back 
\protect\cite{Sch84a}. There is even a report of a very weak single-gamma
branch \protect\cite{Zhe90a}, made possible by coupling to the
atomic shell electrons},
emitting \pn -pairs with sharp sum energy given by 
the transition energy minus the equivalent of the pair's mass. 
The sum energy 
is thus $ \ES=1760$\,keV$-1022$\,keV\,=\,738\,keV. 
This energy can be shared arbitrarily between 
\n\ and \p, leading to
 a broad \ED\ distribution. Due to final state Coulomb 
interaction with the nucleus' field this distribution is shifted to higher \p\
energies and thus positive \ED\ values. Other decay modes of this
transition have been studied in detail theoretically
\cite{Bay59a,Hof89a,Hof94a,Hof95a,Hua56a} as well as experimentally
\cite{Bjo59a,Gre56a,Sch84a,Zhe90a}.

The 1.76\,MeV transition decays to $\approx$30\% by IPC and is populated with
a probability of $1.1\times  10^{-4}$ \cite{Led78a} by the $\beta^-$ decay 
of
\Y with a Q-Value of  2.6\,MeV. Since this isotope is rather short-lived
(64\,h), we used a \Sr\ source, where \Y\ is produced in equilibrium by 
Strontium $\beta^-$ decay with a Q-value of 586\,keV. Thus every \pn\ pair 
created via IPC decay of the first excited state of \Zr\ is accompanied by 
about $2\times 10^5$ electrons from $\beta^-$ decay. Especially the electrons
from the \Sr\ decay cover exactly the range of investigated \p\ and \n\ 
energies of $\approx$150\dots600 keV, determined by the low energy cutoff below
150\,keV of our trigger electronics. To achieve satisfactory counting rates 
for a high statistics experiment, a very strong (in the order of 1\,mCi)  \Sr\
source had to be used, leading to total countrates on the \n\ PAGODA of up to
about 2\,MHz as in the case of heavy-ion experiments. A very thin source had
to be produced in order to keep broadening of the angular correlation by
small-angle scattering of the outgoing leptons as small as possible. A source 
with 1.5$\mu$m Mylar foil plus
a 0.2$\mu$m Aluminum (needed to keep the source
at ground potential) was produced from commercially available \Sr\ 
solution\footnote{Our thanks to Dr. Br\"uchle and Co-workers, 
GSI Nuclear Chemistry Division, for the production of this 
source}.

Using this source we collected a number of 1.3$\times10^5$ \pn -pairs 
in the sharp sum
energy line of the IPC transition over a background of mainly chance
coincidences of \p\ from IPC and $\beta^-$ (Fig. 6a). 
However, the chance coincidence
background can be determined quantitatively and subtraction yields the pure
IPC spectrum (Fig. 6b). This spectrum is essentially background-free with a
ratio of counts in the line (FWHM 15.5\,keV) to the rest of the spectrum
of 5.7:1. The tail on the low energy side is due to a small part of not
suppressed backscattered leptons on either side.

The high statistics in this experiment allows us to subdivide the data into 17
\ED\ intervals of 66\,keV width and the ten opening angle bins of our
spectrometer. 
Theory predicts for this E0 transition an angular correlation of
the form
\begin{equation}\label{anisotropy}
I(\Theta_{e^+e^-}) = a \cdot (1 + \epsilon \cos \Theta_{e^+e^-}),
\end{equation}
where $a$ is given by the conversion coefficient, source strength, detection
efficiency and duration of the measurement.
To determine the angular correlation coefficient $\epsilon$ the above
given
function is fitted to the measured data for each \ED\ bin (Fig 7a),
multiplying the first and last bin by two to correct for the smaller solid
angle (rf. section \ref{setup}). 
However, the measured angular correlation includes also some small-angle
scattering of the leptons in the source. 
To account for this the theoretical results were folded
with small-angle scattering and the experimental response via a Monte Carlo
simulation. 
The result of the simulation was fitted in the same manner as the
measured data (Fig. 7b). 
Now the two values for $\epsilon$ can be compared.
The effective source thickness of 550\mucm\ was calculated from the known 
thickness of the foils and with the assumption that the source material is 
pure. 
However, since the purity  and the mean size of
the crystallites of the active material are not well known, only a upper limit
of about 1500\mucm\ can be given.

In Figure 8 we compare the anisotropy coefficients $\epsilon$ determined 
from the measurement with the theoretical calculations adapted to 
our experimental conditions via Monte Carlo simulation for
various effective source thicknesses indicated.
Even for a  1500\mucm\ thick source the theoretical values lie above the
measured values except for very large $|\ED|$ values. 
This is a hint that the real 
anisotropy
coefficient might be somewhat smaller ($\approx$9\%) than the calculated 
values. However,
within the uncertainties of the current analysis (dominated by the uncertainty
of the source thickness) theory and experiment are 
consistent. The
anisotropy given by older Born approximation results [with
$\epsilon(\ED=0)=1$] is inconsistent with the measurement. 

\subsection{The 1.77\,MeV M1 transition in \Pbs}

Using a rather weak (only a few $\mu$Ci) \Bis\ calibration source~\footnote
{No \Bis\ source material was commercially available to
produce a much stronger source in the same way as with \protect\Sr }
we were also 
able to measure \pn-pairs
from the weak IPC branch of the 1.77\,MeV M1 transition in
\Pbs, leading to a narrow \pn--sum-energy line at $\ES=755\,keV$.
Since we detected here only 
a few pairs per hour, the statistics of this 
measurement does
not allow to subdivide into several \ED\ bins. 
Due to the low rates practically no chance
coincidences occur in this measurement, and the measured \pn\ sum energy
spectrum shown in Fig. 9 is again of the same quality as 
that of Fig. 6b.

However, even the low statistics of this measurement allows us 
to subdivide into the ten opening angle bins and determine the angular
correlation of the pairs. The theoretical prediction for
$I(\Theta_{e^+e^-})$ is
 in this case more complex, but the form given in 
eq.~\ref{anisotropy} 
is the first order in a $\cos(\Theta_{e^+e^-})$ expansion of 
the correct result, so again the same fitting procedure can be used. 
As can be seen in
Fig. 10, the situation is very different from the E0 case reported in the last
section: while for the E0 case the intensity decreases for larger
$\Theta_{e^+e^-}$ values, in this case the  
intensity increases in the $\Theta_{e^+e^-}$ range covered by our
setup, thus yielding a negative value of $\epsilon = -0.12\pm0.06$ for the
anisotropy coefficient. 
This is in strong disagreement with the
theoretical results derived using the Born approximation, but in good
agreement with the most recent theoretical results (see Fig. 1).

These new results have some impact on estimates for the expected strength of the
accompanying $\gamma$ line, if some of the previously reported sum energy
lines found in heavy ion experiments could be explained by IPC processes. 
According to these findings the 
pair efficiency of our setup is much higher for angular correlations 
having the pattern of the M1 case than for angular correlations derived 
from Born approximation. 
This is particularly 
true for lines appearing only in the 180\degree\ bin of the setup,
where the Doppler shifts of the two leptons almost cancel:
For this specific situation the detection efficiency  is an order of magnitude
larger for $\epsilon=-0.12$ than for $\epsilon=1$. 
We would thus expect a factor of ten weaker $\gamma$ 
line for $\epsilon=-0.12$  
than in the old estimate~\cite{Koe93a}. 
Only a detailed re-analysis of the old data~\cite{Koe93a}
can clarify whether IPC could be
the origin of these weak lines.
It should be noted in this context that
recent results, obtained from high-resolution Doppler-shift spectroscopy
in heavy-ion collisions at the Coulomb barrier,
have indeed revealed that weak \pn--sum-energy lines from IPC transitions
of moving emitters can appear in the measured spectra~\cite{Hei97a}.

\section{Summary}

The measurements presented in this work prove the capability of our setup to
detect the sharp \pn\ sum energy line of the IPC decay of excited heavy nuclei
despite the several orders of magnitude more intense background of electrons
and $\gamma$ rays from the same source. The measurement with the \Sr\ source
clearly shows that 
this is even the case for very high count rates in the
MHz range on the electron detectors, as they are typical also for in-beam
measurements (see e.g. Ref.~\cite{Lei97a}). 

The results of the first energy and angle resolved measurements of IPC pairs
---made more than 60 years after the first theoretical prediction of this decay
branch--- prove that Born approximation 
is not suitable for the treatment of
IPC transitions of only a few times the threshold in heavy nuclei even for
the most simple E0 case and by no means for magnetic multipole transitions.
The theoretical results for the angular correlation presented here, taking into
account all relativistic corrections, the finite size of the nucleus and also
the magnetic substates for magnetic transitions, are in agreement with the
experimental results shown in this paper. 
The combined theoretical and experimental
efforts of our groups have thus led to a better understanding of a fundamental
decay channel of excited nuclei that was assumed to be very well understood
for decades.


\newpage
\begin{center}
\subsection*{\bf TABLES}
\end{center}
\begin{table}[ht]
\caption{\pn\ opening angle bins of the ORANGE setup}
\vspace*{5mm}
\begin{tabular}{*{10}{c|}c}
 $\Delta\phi\,[^\circ]$ & 0 & 20 & 40 & 60 & 80 & 100& 120 & 140 & 160 & 180
 \\\hline
 $\langle\Theta_{e^+e^-}\rangle\,[^\circ]$     & 70 & 73 & 80 & 90 & 102& 116  
 & 131 & 145 & 159 & 167 \\\hline 
 $\Delta\Theta_{e^+e^-}\,[^\circ]$ &$\pm$14&$\pm$13&$\pm$13&$\pm$12&$\pm$11&
 $\pm$9 & $\pm$9  &$\pm$8  & $\pm$8  & $\pm$7 \\\hline
 $d\Omega^2\,[sr^2]$ & 0.31 & 0.62 & 0.62 & 0.62 & 0.62& 0.62 & 0.62 & 0.62 
& 0.62 & 0.31 

\end{tabular}
\end{table}

\newpage
\begin{center}
\subsection*{\bf FIGURE CAPTIONS}
\end{center}

\vspace*{1cm}

{\bf Fig.1:} Doubly differential pair conversion coefficient plotted versus
the opening angle $\Theta$ of the electron-positron pair for an M1 transition
in $^{207}$Pb. The transition energy amounts to 1.77\,MeV, the kinetic
positron energy was fixed to 450\,keV.
The dotted line represents the Born approximation result, which
is valid for small nuclear charge numbers. In contrast to this, 
the DWBA calculation yields a 
nearly isotropic distribution. One can also deduce the influence of
the nuclear size (full line) which is compared to the point nucleus result 
(dashed line).

\vspace*{1.0cm}
{\bf Fig.2:} Angular correlation for a M1 transition with transition
energy 1.77\,MeV. The positron energy is fixed to 370\,keV (symmetric
splitting of the nuclear transition energy on electron and positron).
In order to study the effect of the nuclear charge the angular distribution
is plotted for the nuclear charge numbers $Z=92$ (full curve), $Z=82$
(dashed curve), $Z=40$ (dash-dotted line), as well as for the Born 
approximation result (dotted line).

\vspace*{1.0cm}
{\bf Fig.3:}
Anisotropy coefficient $\epsilon$ plotted versus the energy difference \ED\
for E0 conversion assuming a transition energy of 1.76\,MeV and
three nuclear charge numbers $Z=0$ (Born approximation result, dashed line),
$Z=40$ (full line) and $Z=92$ (dotted line). The $Z=40$ curve is the
prediction of the anisotropy coefficient for the E0 transition in\Zr.
 
\vspace*{1.0cm}
{\bf Fig.4:}
Schematic view of the new double-Orange setup. 
Each of the $\beta$-spectrometers is
equipped with a $\beta$-multidetector system of 72 Si (PIN) diodes
(e$^{\pm}$--Pagodas).
The forward spectrometer is
surrounded by 18 position-sensitive heavy-ion detectors (PPAC), and
contains a further PPAC detector in its center.
Also shown is the rotating target wheel and the Ge(i) $\gamma$-ray detector.
The only change made for the current experiment is that the target wheel
was used in a stationary mode, on a target position of which 
the radioactive sources were mounted.

\vspace*{1.0cm}
{\bf Fig.5.} 
Experimental signature of the detected positrons in the \Zr\ measurement. 
The abscissa shows the normalized difference between the momentum of the 
focused positrons calculated from the pulse height of the Si detectors and 
their
momentum determined from the magnetic field of the spectrometer.
Only events falling in a narrow window centered around zero are 
accepted in the analysis. 
The background is dominated by electrons from the five orders of
magnitude more intense $\beta^-$ rays, reaching the positron detectors after
multiple scattering in the spectrometer. 

\vspace*{1.0cm}
{\bf Fig.6:}
e$^+$e$^-$-sum-energy spectrum from the \Sr\ source. Left part: raw spectrum,
overlayed with the chance
 coincidence background due to the high
(1\ldots2\,MHz) $\beta^-$ count rate determined from not prompt coincidences.
Right part: Sum energy spectrum after subtraction of chance coincidences. 
The small
low energy tail is due to a small fraction of backscattered leptons not
rejected by the lepton identification (see text). The sum energy line from the
1.76\,MeV IPC decay in \Zr\ is measured essentially
background-free despite the five orders of magnitude higher $\beta^-$
background from the decay of \Sr\ and \Zr.
 The resolution is 15.5\,keV (FWHM),
the total content of the spectrum 1.59$\times$$10^5$

\vspace*{1.0cm}
{\bf Fig.7:}
An illustrative example of the determination of the angular correlation 
coefficient $\epsilon$ for
both, measurement (left part) and theory (right part), which are treated 
in the same manner:
The function 
$I = a (1+\epsilon\cos\Theta_{e^+e^-})$ 
is fitted to the number of
counts in the angular bins of the ORANGE setup at an energy difference
centered at zero.
The theoretical result is gained by folding the theoretical prediction 
with small-angle 
scattering in the source and the acceptance of the 
ORANGE setup by a Monte Carlo simulation.
The error quoted in the experimental value of 
$\epsilon=0.83\pm0.03$
is solely of statistical origin. 

\vspace*{1.0cm}
{\bf Fig.8:}
The extracted  angular correlation  coefficient
$\epsilon$ for IPC decay of the 1.76\,MeV
transition in \Zr{} as a function of the energy difference \ED{} of the
emitted \pn--pairs. 
The theoretical prediction (solid line) was folded with
small-angle scattering in a source with a thickness 
of 750\mucm\ (dashed line) and 1500\mucm\
(dotted line). The points show the experimental result. 

\vspace*{1.0cm}
{\bf Fig.9:}
\pn\ sum energy spectrum measured with a weak \Bis\ source. Due to the low
count rates, 
chance coincidences of leptons from IC and $\beta^+$ decay
branches are negligible. Again the sharp sum energy peak from the IPC decay of
the 1.77\,MeV M1 transition in \Pbs\ is measured essentially background-free,
even though a huge (five orders of magnitude larger) background of $\gamma$'s,
\n\ from IC and $\beta^+$ is present.

\vspace*{1.0cm}
{\bf Fig.10:}
Measured angular distribution of the IPC line from the 1.77\,MeV M1 
transition in \Pbs.
Again the intensity in the first and last bin have been multiplied
by two to
account for their smaller solid angle. A fit done in the same manner as in
Fig.~7 yields a negative anisotropy coefficient $\epsilon=-0.12\pm0.06$, i.e. 
the minimum of
the intensity is at 90\degree\ rather than 180\degree. 
%
%
%
%
\newpage
\pagestyle{empty}  

\vspace*{3cm}
\begin{center}
\epsfig{file=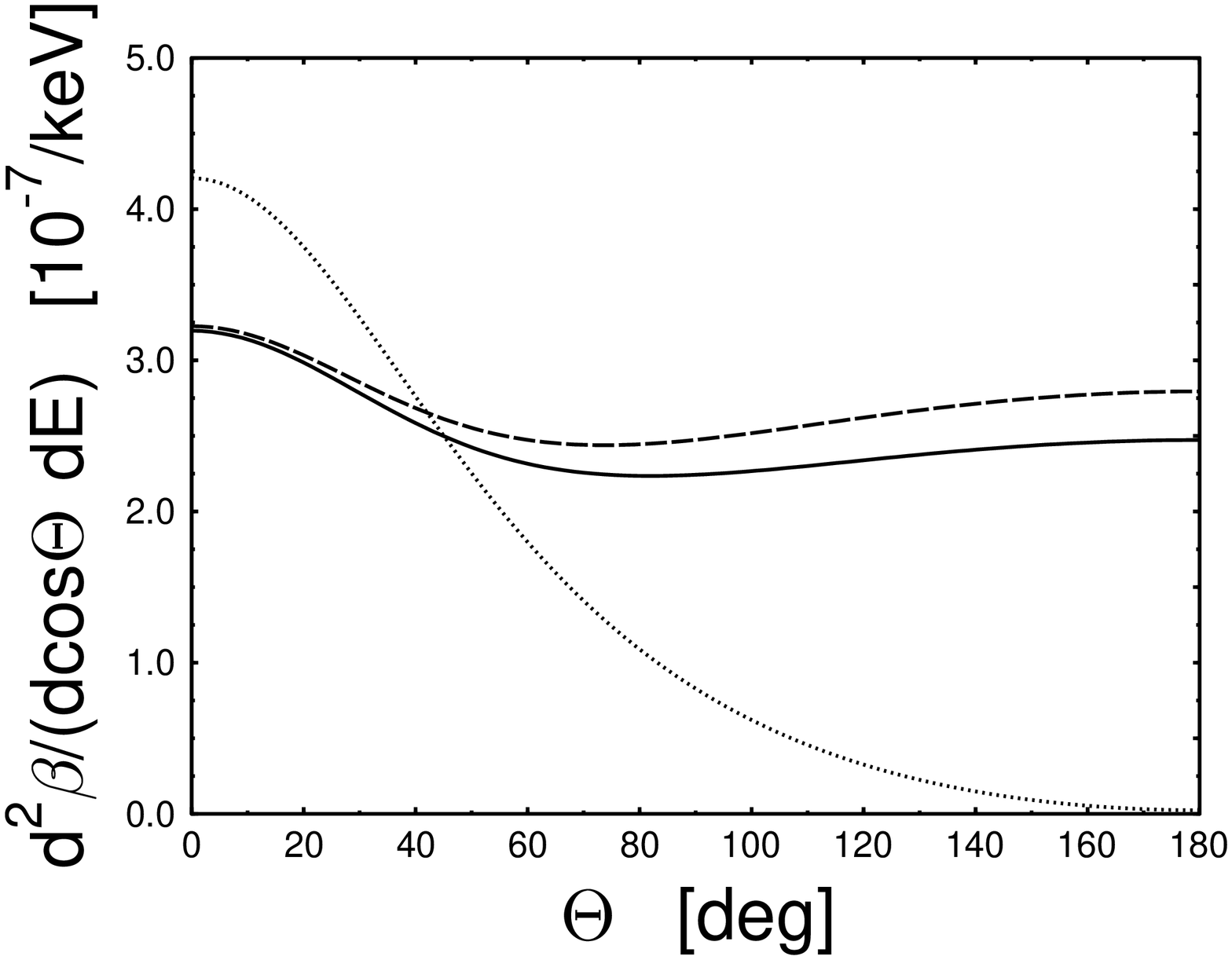,width=\singlepicwidth}
\end{center}

 \vspace*{6cm}
 {\Large \bf Figure 1} 

\vspace*{5 mm}
(U. Leinberger {\em et al.}, Zeitschrift f\"ur Physik A)
 
\newpage

\vspace*{3cm}
\begin{center}
\epsfig{file=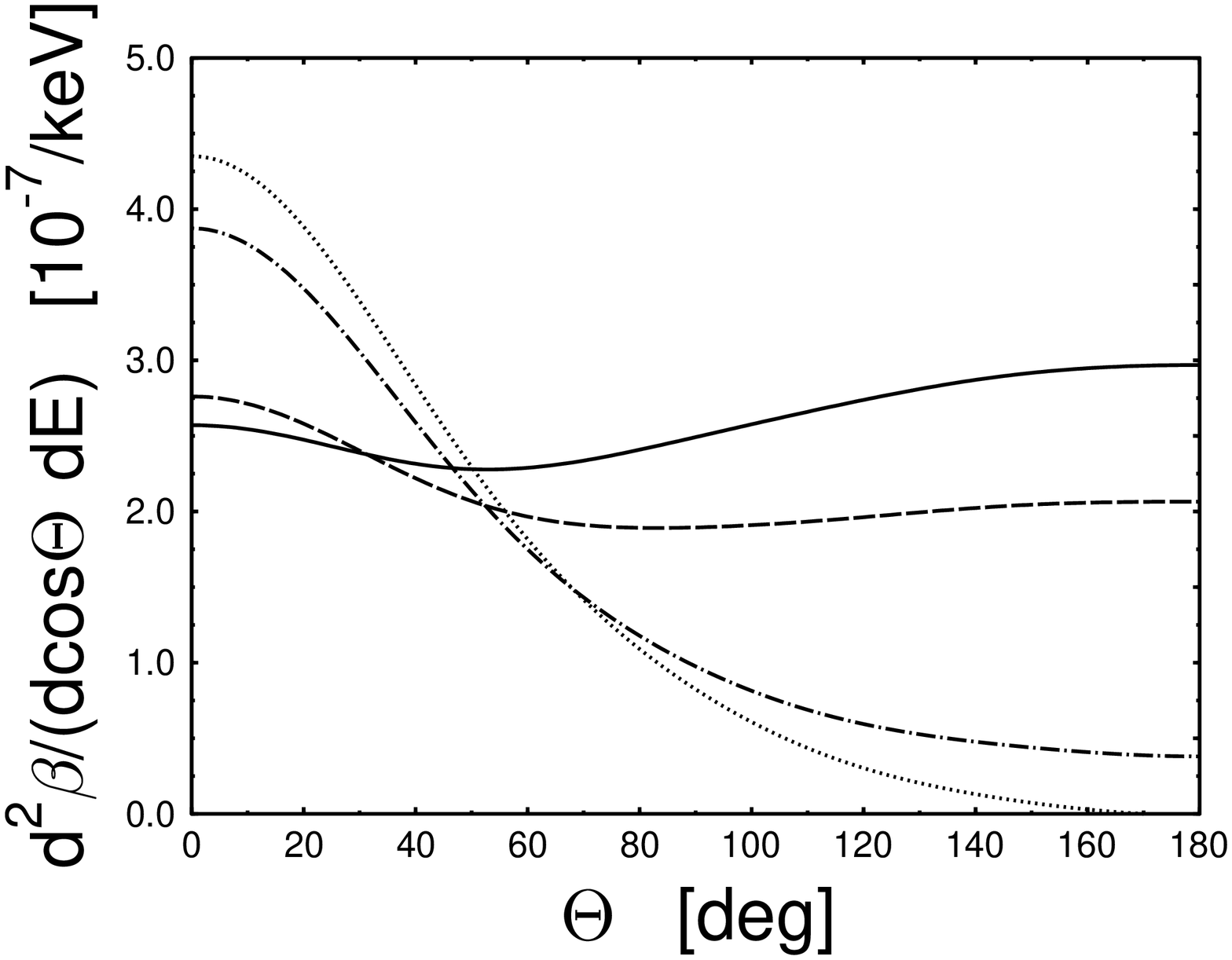,width=\singlepicwidth}
\end{center}

\vspace*{3cm}
 {\Large \bf Figure 2} 

\vspace*{5 mm}
(U. Leinberger {\em et al.}, Zeitschrift f\"ur Physik A)

\newpage

\vspace*{3cm}
\begin{center}
\epsfig{file=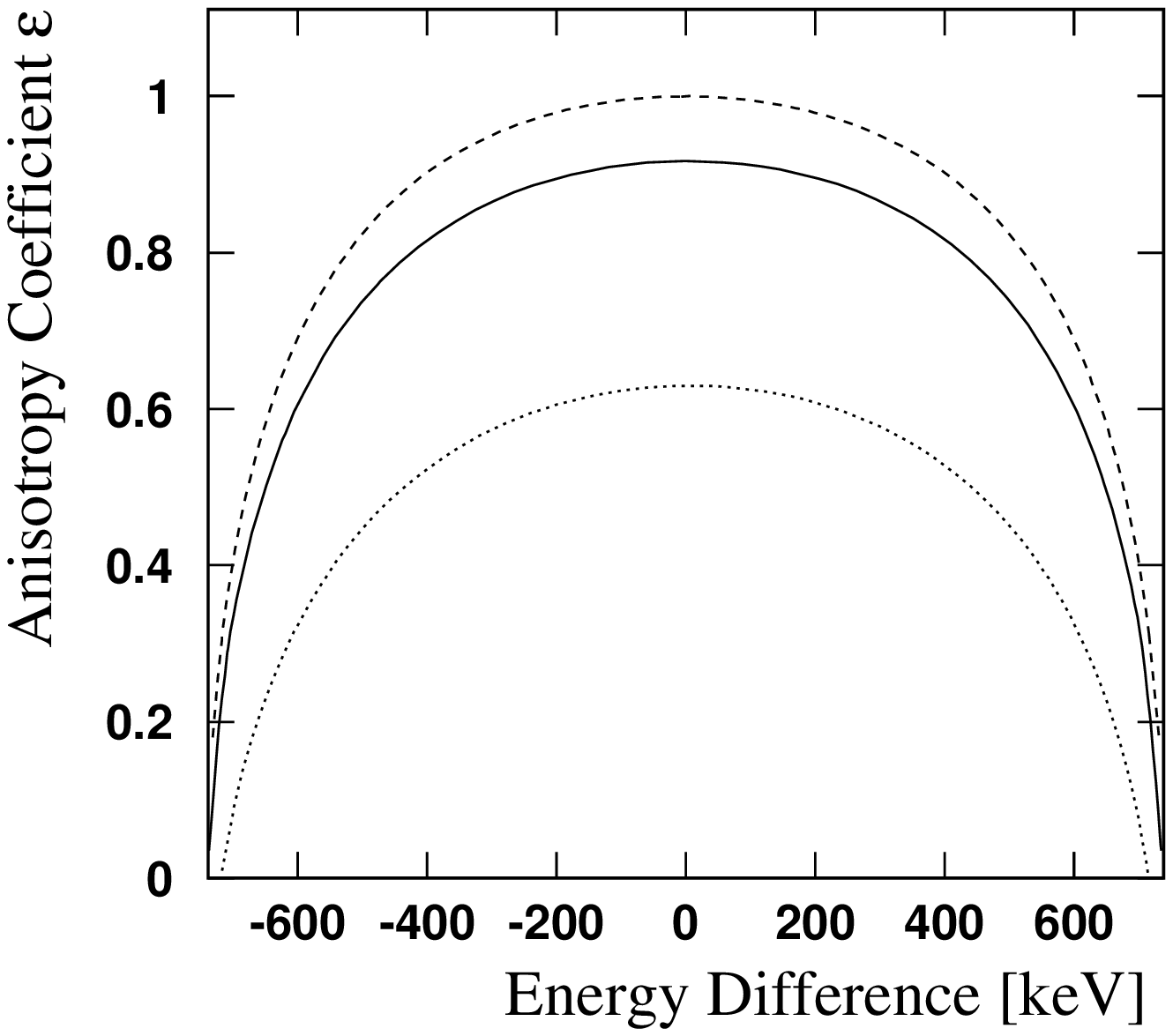,width=\singlepicwidth}
\end{center}

\vspace*{3cm}
 {\Large \bf Figure 3} 

\vspace*{5 mm}
(U. Leinberger {\em et al.}, Zeitschrift f\"ur Physik A)

\newpage
\pagestyle{empty}  

\vspace*{3cm}
\begin{center}
\epsfig{file=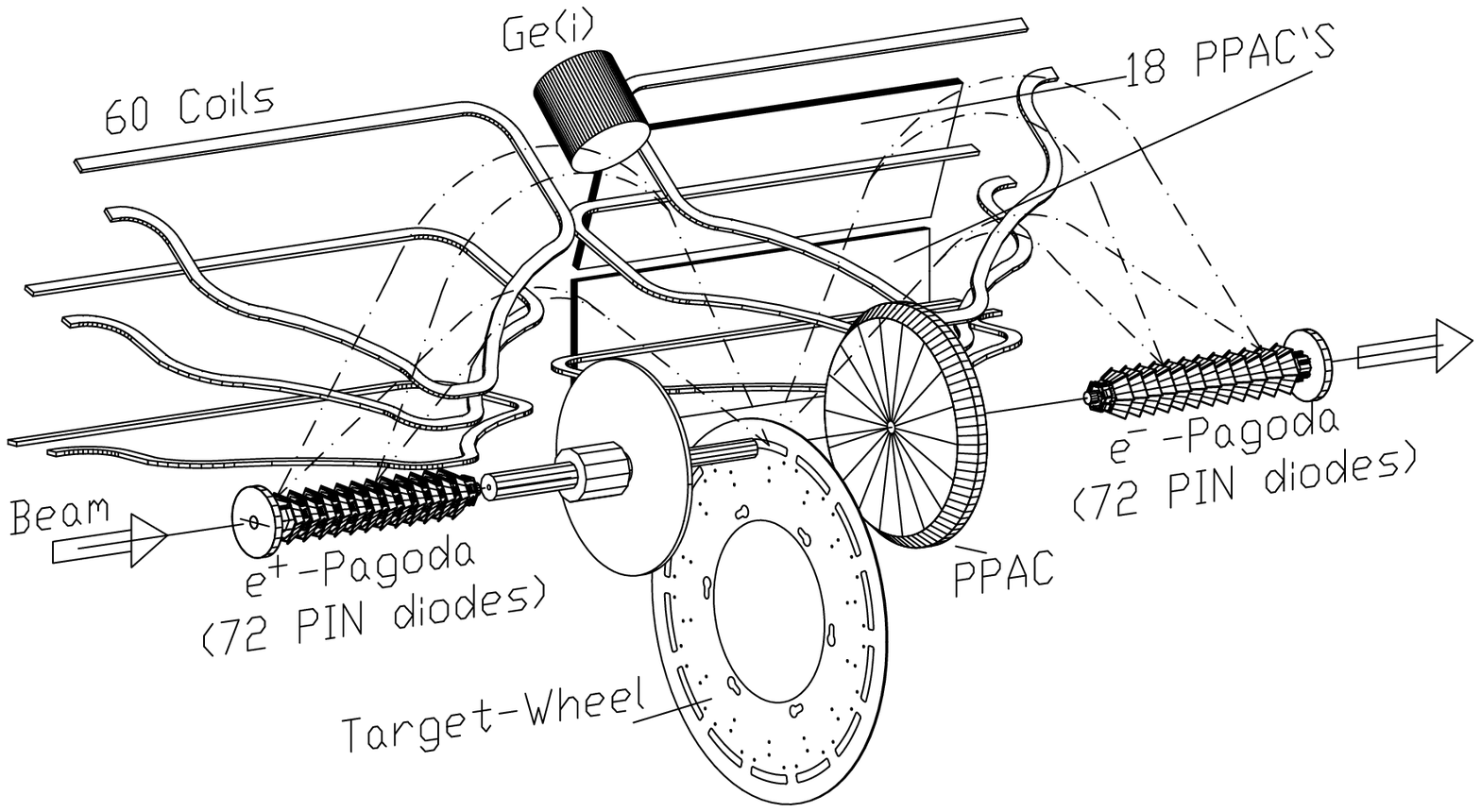,width=1.1\linewidth}
\end{center}

 \vspace*{6cm}
 {\Large \bf Figure 4} 

\vspace*{5 mm}
(U. Leinberger {\em et al.}, Zeitschrift f\"ur Physik A)
 
\newpage

\vspace*{3cm}
\begin{center}
\epsfig{file=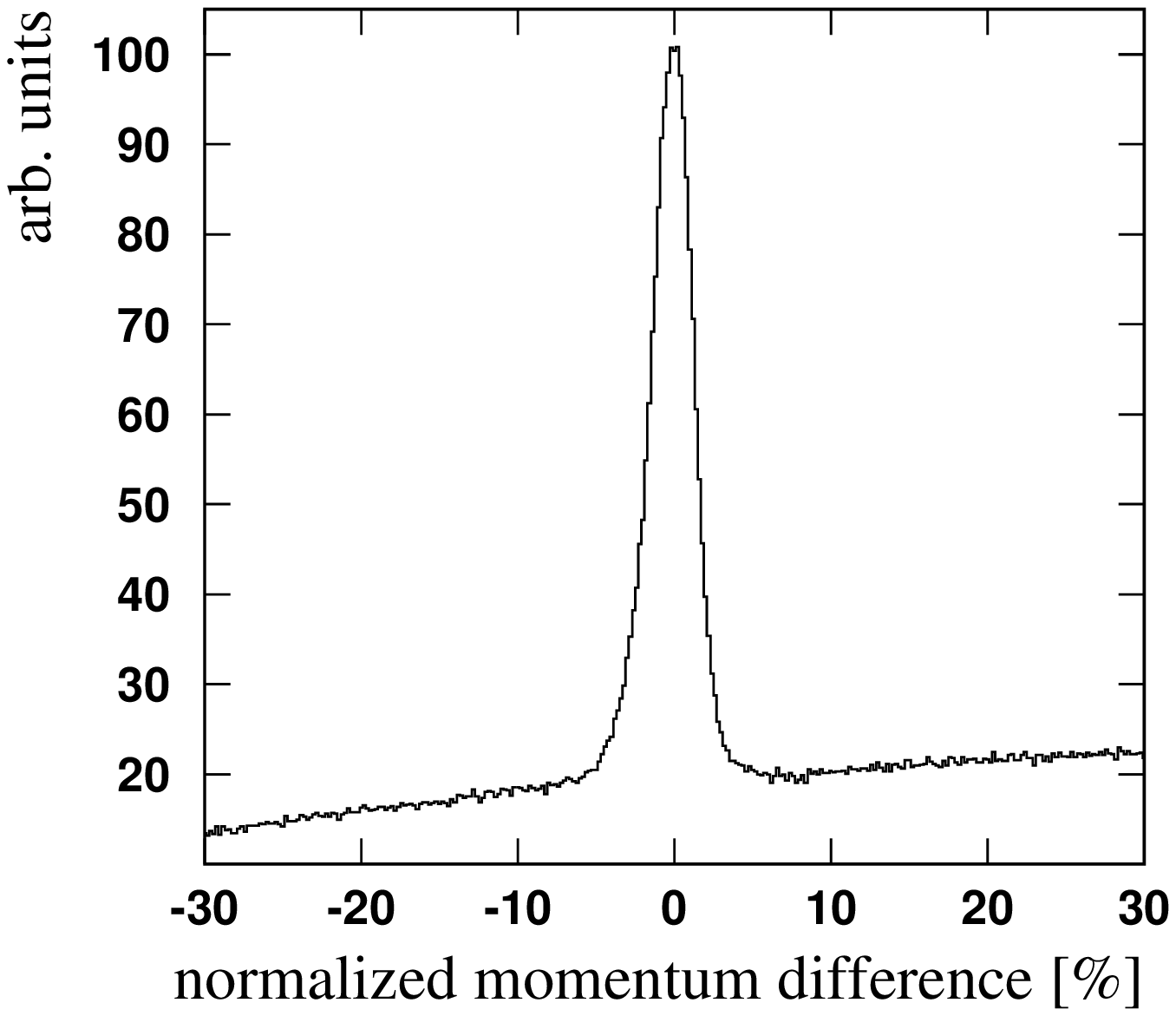,width=\singlepicwidth}
\end{center}

\vspace*{3cm}
 {\Large \bf Figure 5} 

\vspace*{5 mm}
(U. Leinberger {\em et al.}, Zeitschrift f\"ur Physik A)

\newpage

\vspace*{3cm}
\hspace*{\firstpicshift}\epsfig{file=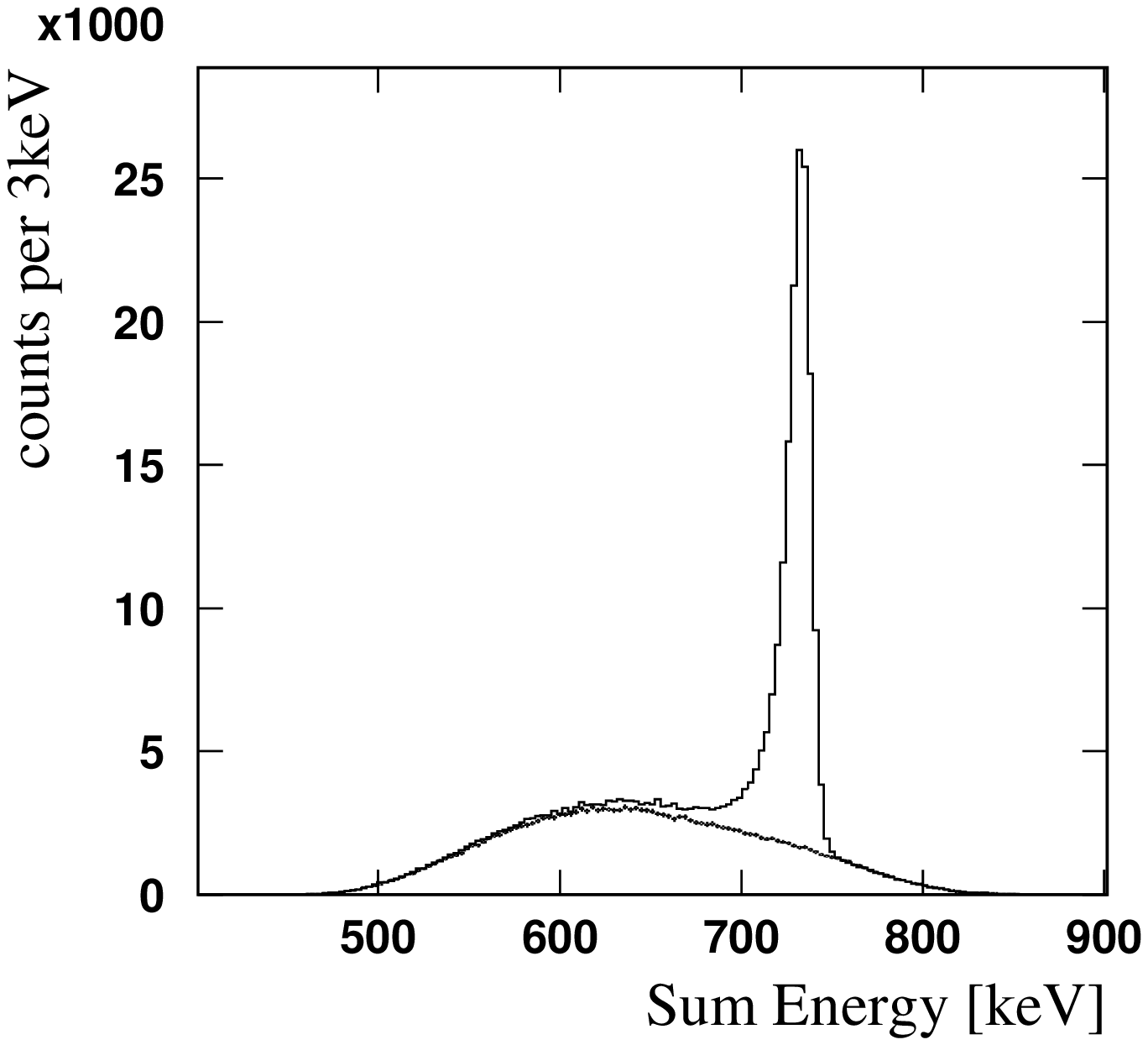,width=\doublepicwidth}
\hspace*{\secondpicshift}\epsfig{file=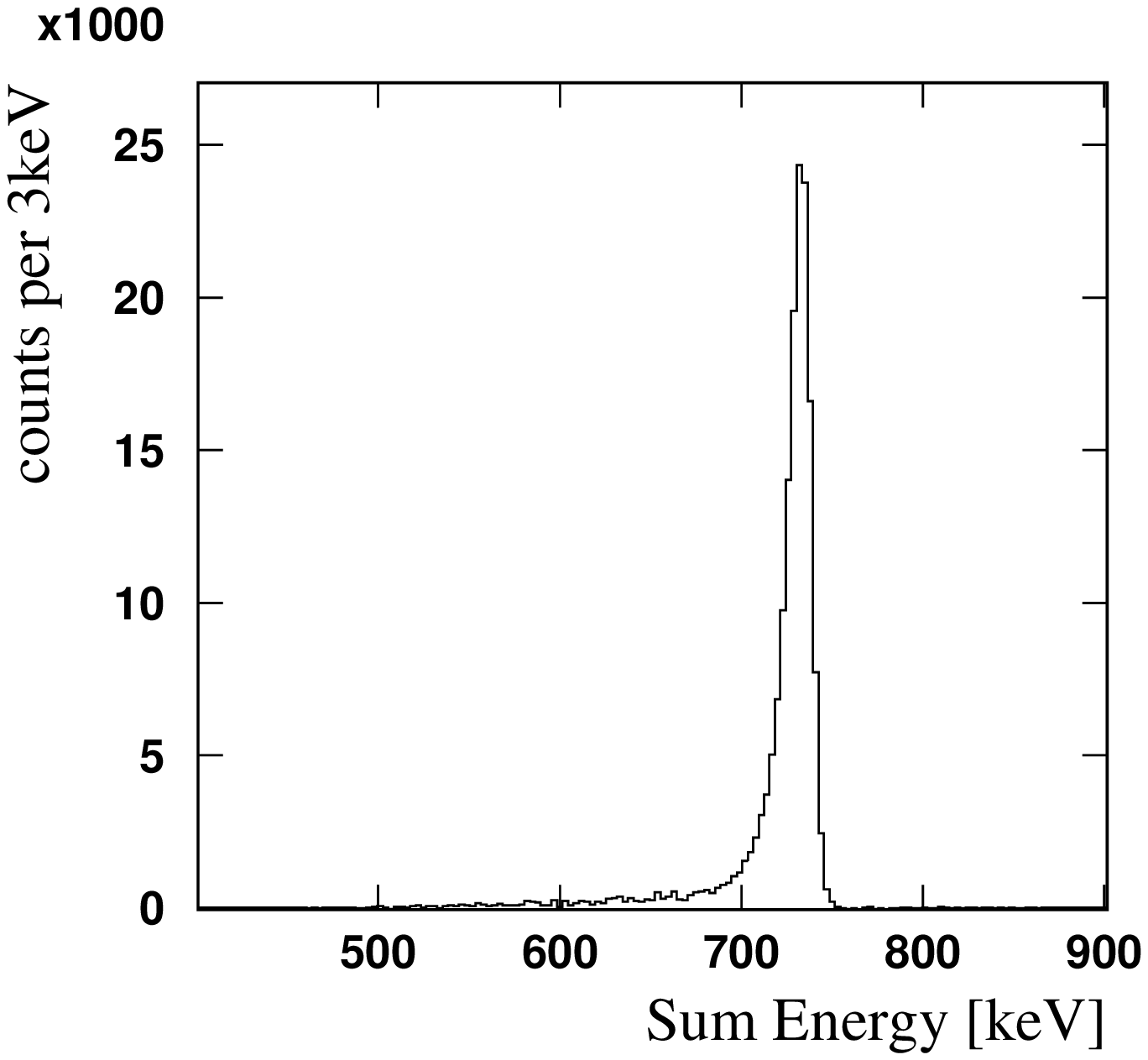,width=\doublepicwidth}

 \vspace*{3cm}
 {\Large \bf Figure 6} 

\vspace*{5 mm}
(U. Leinberger {\em et al.}, Zeitschrift f\"ur Physik A)

\newpage

\vspace*{3.0cm}
\hspace*{95pt}Experiment\hfill Theory\hspace*{60pt}\\[-14pt]\vspace*{-59pt}\\
\hspace*{\firstpicshift}\epsfig{file=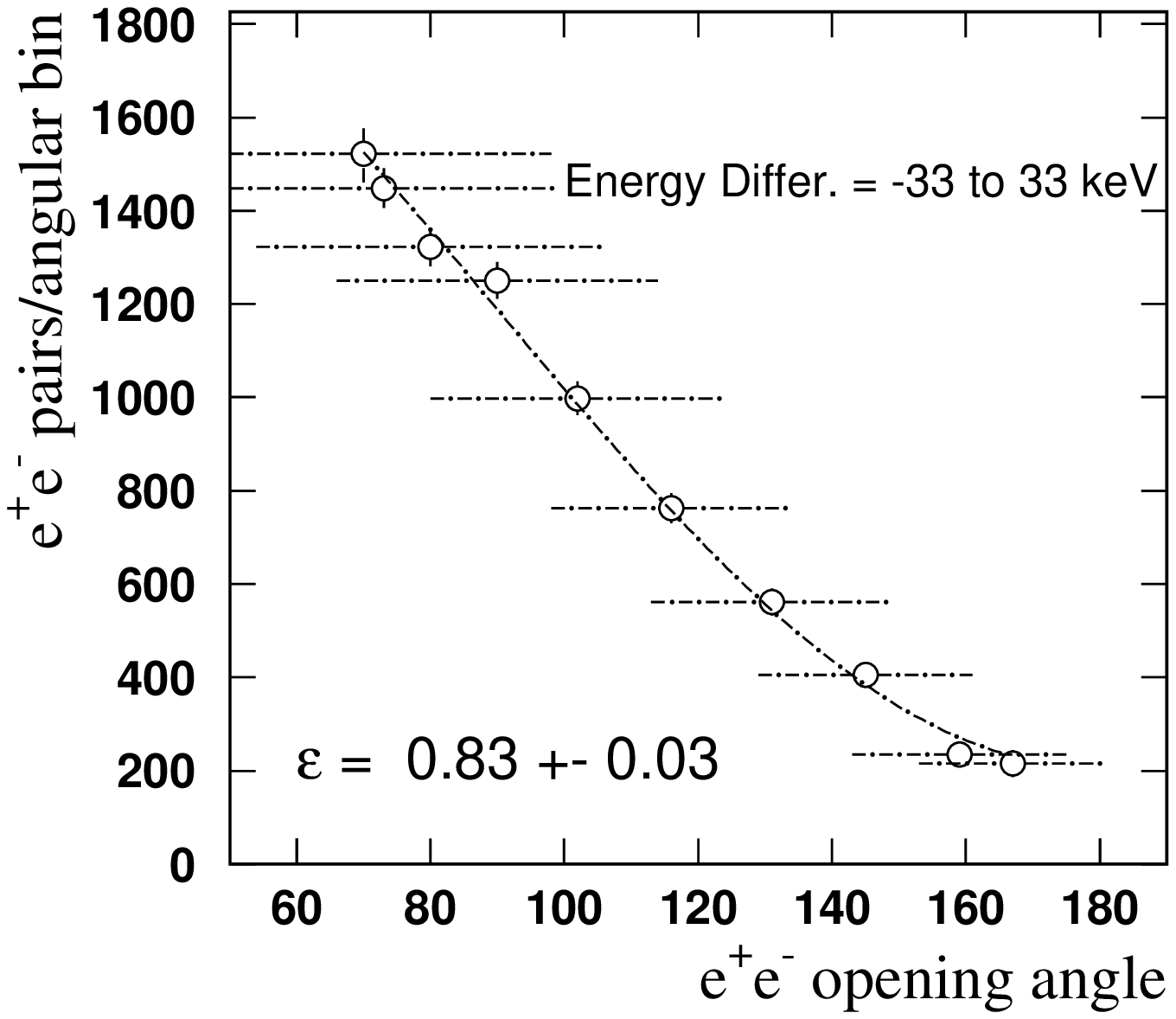,width=\doublepicwidth}
\hspace*{\secondpicshift}\epsfig{file=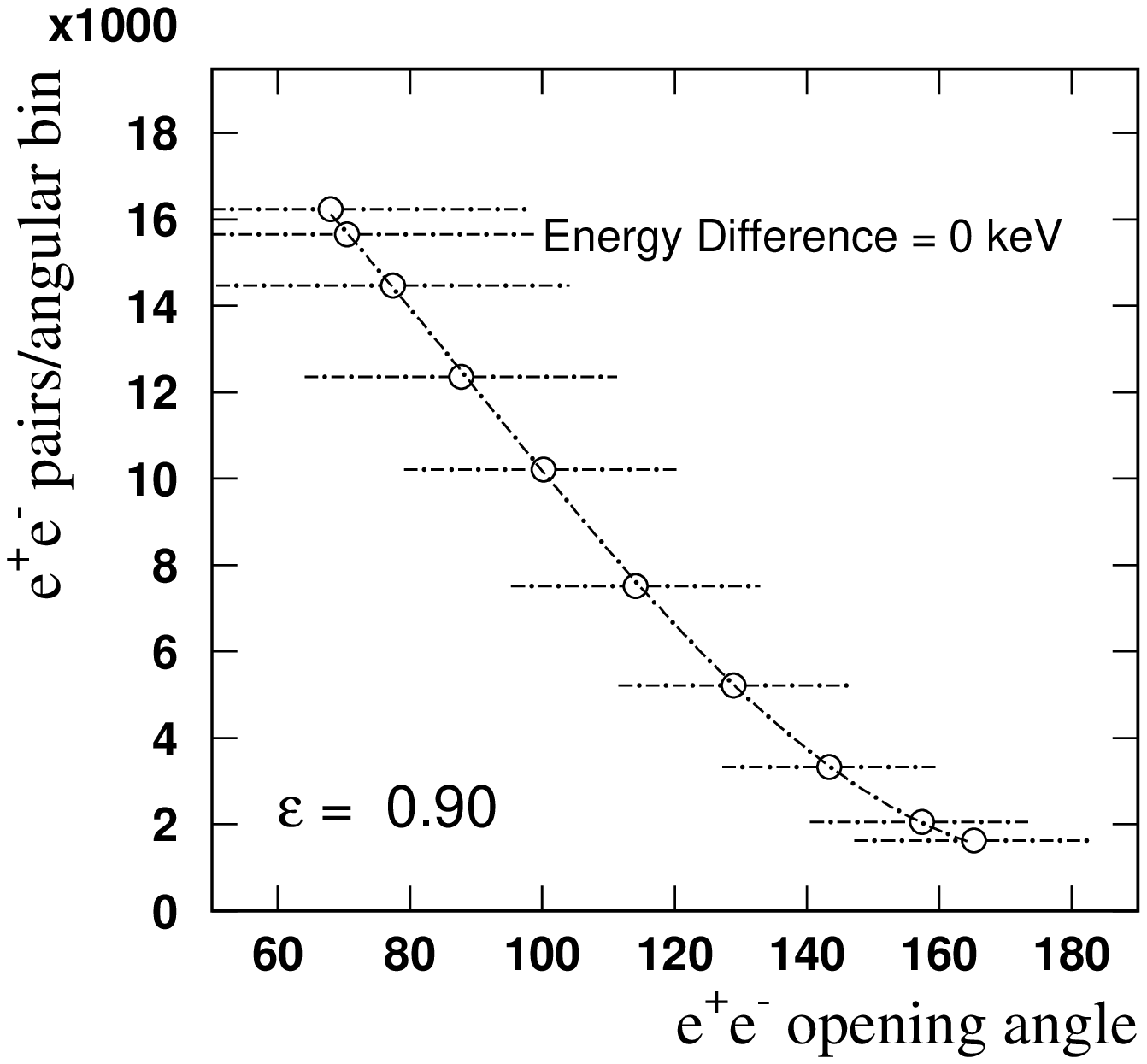,width=\doublepicwidth}

 \vspace*{3cm}
 {\Large \bf Figure 7} 

\vspace*{5 mm}
(U. Leinberger {\em et al.}, Zeitschrift f\"ur Physik A)

\newpage

\vspace*{3cm}
\begin{center}
\epsfig{file=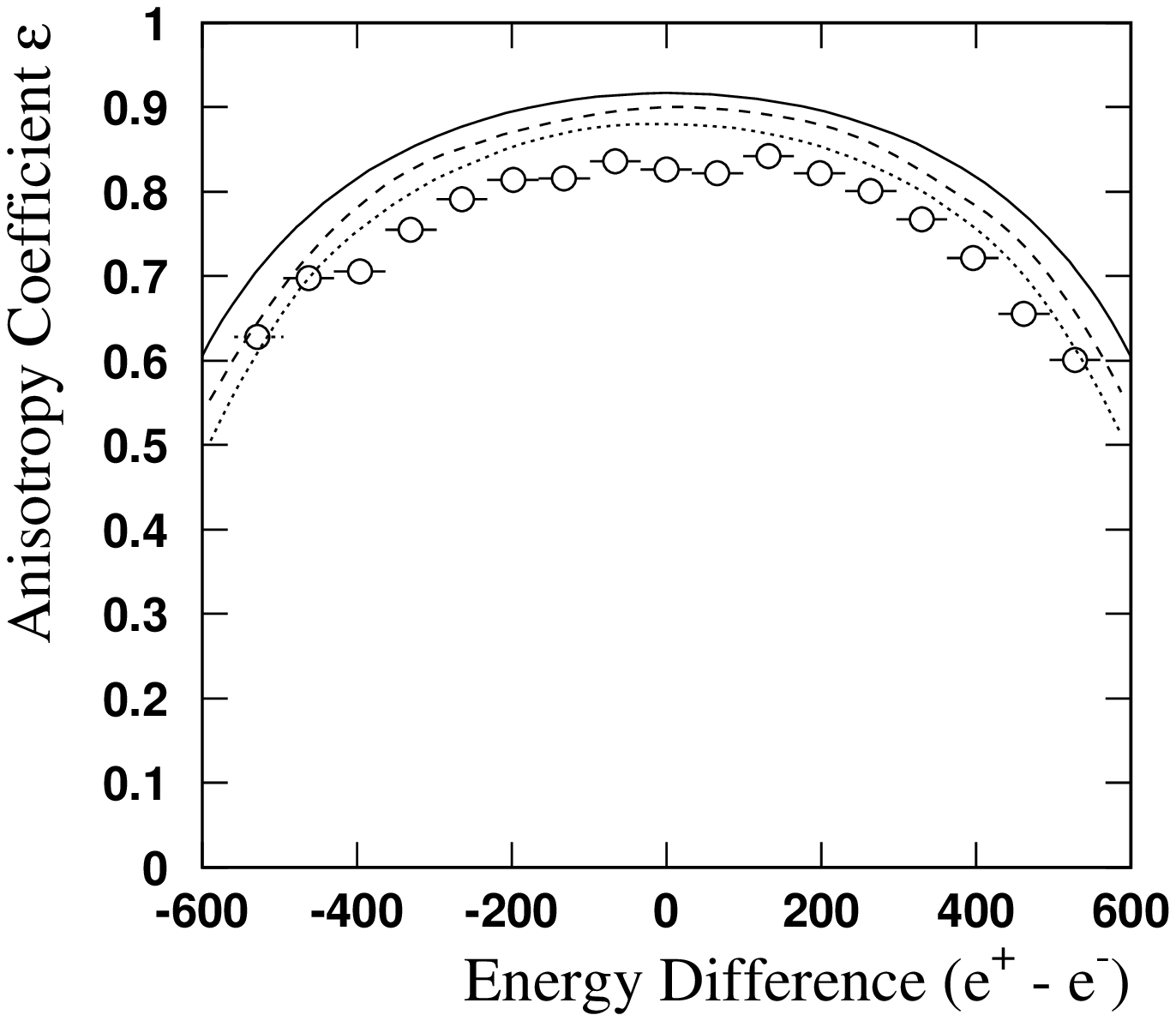,width=\singlepicwidth}
\end{center}

\vspace*{3cm}
 {\Large \bf Figure 8} 

\vspace*{5 mm}
(U. Leinberger {\em et al.}, Zeitschrift f\"ur Physik A)

\newpage

\vspace*{3cm}
\begin{center}
\epsfig{file=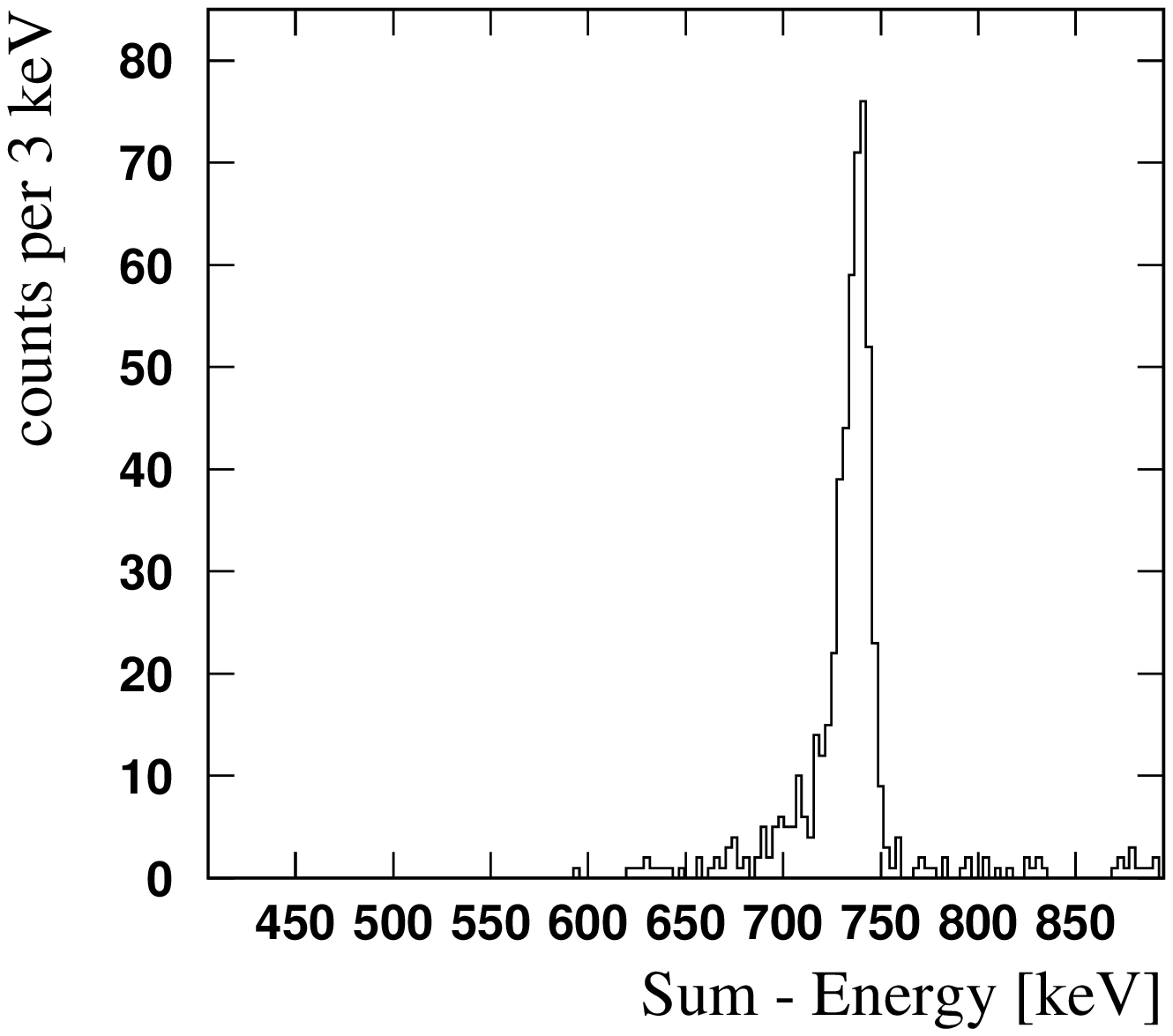,width=\singlepicwidth}
\end{center}

\vspace*{3cm}

 {\Large \bf Figure 9} 

\vspace*{5 mm}
(U. Leinberger {\em et al.}, Zeitschrift f\"ur Physik A)
\newpage

\vspace*{3cm}
\begin{center}
\epsfig{file=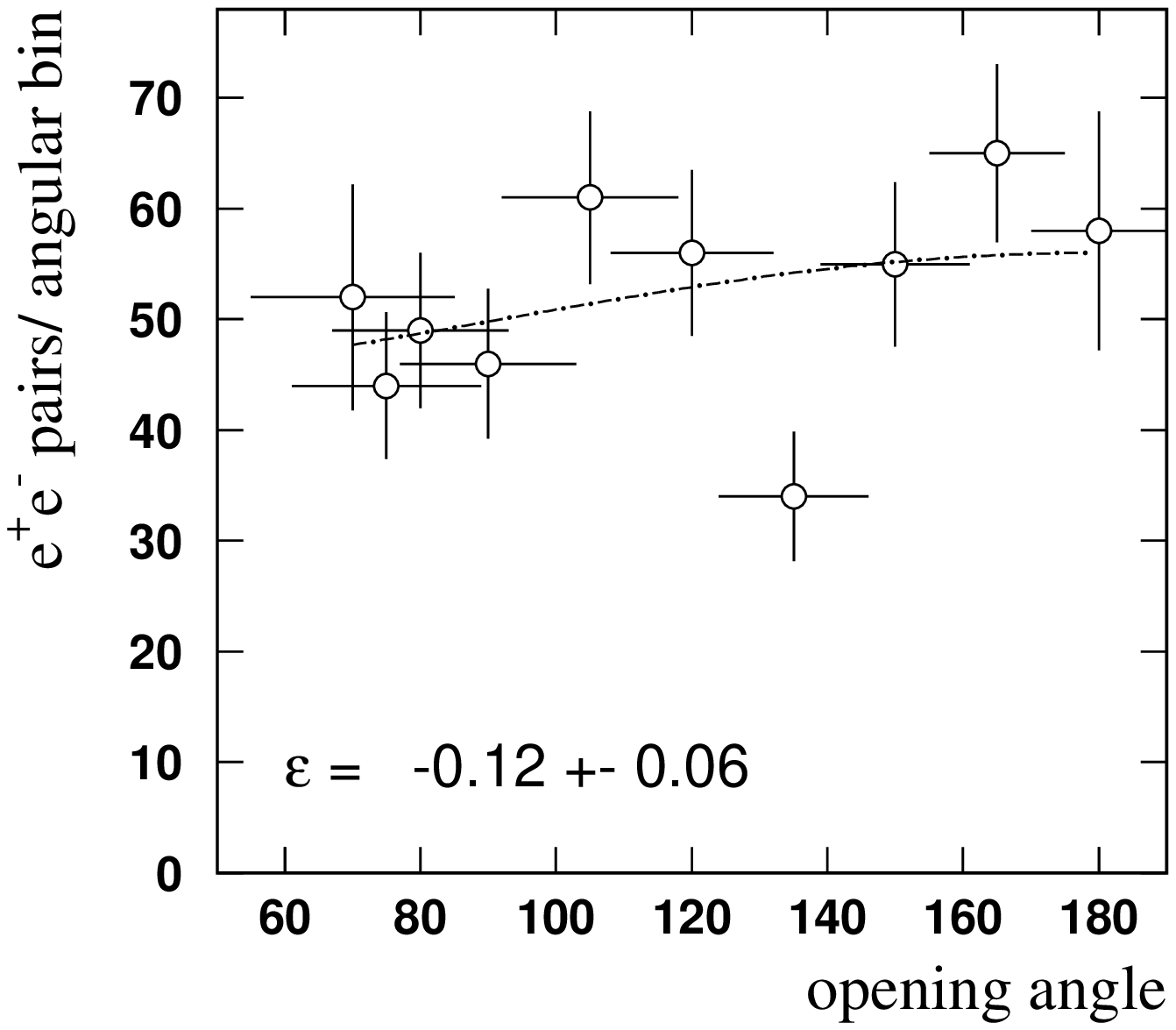,width=\singlepicwidth}
\end{center}

\vspace*{3cm}
 {\Large \bf Figure 10} 

\vspace*{5 mm}
(U. Leinberger {\em et al.}, Zeitschrift f\"ur Physik A)

\end{document}